\documentclass{article}
\usepackage[margin = 1in]{geometry}
\usepackage{amsmath}
\usepackage{amssymb}
\usepackage{gensymb}

\usepackage{multirow}
\usepackage{graphicx}
\usepackage{hyperref}
\usepackage{subfig}
\title{}
\usepackage{authblk}

\title{Quantum Entanglement in Top Quark Pair Production}

\author{Mira Varma}
\author{Oliver K. Baker\thanks{Corresponding author: \texttt{oliver.baker@yale.edu}}}

\affil{Department of Physics, Yale University, New Haven, CT 06520}

\date{}

\begin{document}
\maketitle

\begin{abstract}
Top quarks, the most massive particles in the standard model, attract considerable attention since they decay before hadronizing. This presents physicists with a unique opportunity to directly investigate their properties. In this letter, we expand upon the work of G. Iskander, J. Pan, M. Tyler, C. Weber and O. K. Baker to demonstrate that even with the most massive fundamental particle, we see the same manifestation of entanglement observed in both electroweak and electromagnetic interactions. We propose that the thermal component resulting from protons colliding into two top quarks emerges from entanglement within the two-proton wave function. The presence of entanglement implies the coexistence of both thermal and hard scattering components in the transverse momentum distribution. We use published ATLAS and CMS results to show that the data exhibits the expected behavior. 

\vspace{0.95cc}
\parbox{24cc}{{\it Key words}: Quantum entanglement, Entanglement entropy, Top physics, Heavy quark production}
\end{abstract}
\section{Introduction}
Prior literature has established that the transverse momentum distribution of hadrons is best described by fitting the sum of an exponential and a power law. (See Refs.\thinspace\cite{Bylinkin:2014vra, Bylinkin:2012bz} for more detail). The power law portion of the fit, which represents hard scattering, is well understood: it arises from the sizable momentum transfer between the quarks and gluons \cite{Bylinkin:2014vra}. The thermal behavior of the transverse momentum distribution, on the other hand,  remains a mystery in particle physics. There have been several competing ideas about why this behavior is present  \cite{Heinz:1999kb}-\cite{Becattini:2004td}. A common belief is that thermalization arises through re-scattering after nuclei collide \cite{Becattini:2008tx}. This explanation is limited: it cannot explain the origin of thermalization in proton-proton ($pp$) collisions. 

A universal explanation for the behavior of the transverse momentum distribution is that it is due to entanglement between parts of the wave functions of the colliding particles. This idea has been studied for several interactions. G. Iskander et al. showed that for weak interactions, specifically neutrino scattering, there is entanglement between the probed and unprobed regions of the nucleon in the collision \cite{Iskander:2020rkb}. This theory has also been studied for electroweak processes, namely, deep inelastic scattering (DIS). K. Zhang et al. calculated the von Neumann entropy (interpreted as the entanglement entropy) of the DIS system, which they proposed was caused by entanglement between the probed and unprobed regions of the proton \cite{Zhang:2021hra}. 

As discussed in Ref.\thinspace\cite{Baker:2017wtt}, when two protons collide, the entire system undergoes a “quench” due to the sudden presence of a collision and a spectator region. The Hamiltonian is evolved, meaning, $H = H_{0} \rightarrow H_{0} + V(t)$, where $V(t)$ is the effect of a pulse of the color field. The uncertainty principle suggests that the momentum transfer of the collision, $Q$, and the proper time (time measured in the particle’s rest frame), $\tau$, are related by: $\tau \sim 1/Q$ \cite{Baker:2017wtt}. If we approximate a $pp$ collision as a short pulse of a (chromo)\thinspace electric field, the effective temperature parameter (arising from thermalization), satisfies the following relation \cite{Baker:2017wtt}: 

\begin{equation}
T_{th} \simeq (2\pi\tau)^{-1} \simeq \frac{Q}{2\pi}.
\end{equation} 

In Eq.\thinspace1, $T_{th}$ is a parameter of the thermal component of the transverse momentum distribution \cite{Baker:2017wtt}. (See Refs.\thinspace\cite{Kharzeev:2005iz, Kharzeev:2006zm, Bylinkin:2014vra} for more detail). 

In this letter, we extend this idea to the most massive known particle in the standard model, the top quark. The top quark is particularly interesting due to its short lifetime (i.e. top quarks decay before hadronizing). We propose that when two protons collide, the thermal part of the transverse momentum distribution is caused by entanglement in the wave function of the proton-proton system. This entanglement is between a collision region, where the two protons overlap, which we call $A$, and a region where the two protons do not overlap, which we call $B$. In other words, we are proposing that when entanglement is present, the transverse momentum distribution has both thermal and hard scattering components. Naturally, therefore, where there is no entanglement, the thermal component is absent.

\section{Background}
In the general quantum information formalism, a pure state of two quantum systems, $A$ and $B$, is denoted as,
\begin{equation}
\rho_{pure} = |\psi\rangle  \langle\psi|,
\end{equation}
where $\rho_{pure}$ is the full density matrix describing the state $|\psi\rangle$. 
Physically, when a system is in a pure state, there is complete information about the wave function. On the other hand, if a system is in a mixed quantum state, there is incomplete information about the wave function at every point in time. A mixed state is a (probabilistic) weighted sum of pure states. 
The full density matrix for a mixed state is: 
\begin{equation}
\rho_{mixed} = \sum_{k=1}^{N} p_{k} |\psi_k\rangle\langle\psi_k|,
\end{equation}
where $p_{k}$ are the weights and $\{|\psi_k\rangle\}$ is a set of pure states. Therefore, if $N=1$, Eq. (3) reduces to a pure state. The reduced density matrix of either one of the subsystems is needed in order to calculate the entanglement entropy. For any state of the two quantum systems, $A$ and $B$, we can describe system $A$ by tracing over system $B$. This results in a reduced density matrix of subsystem $A$, which is expressed as follows:
\begin{equation}
\rho_A =  \sum_{k}\langle k|\rho_{AB}|k \rangle_B= \mathrm{Tr}_B (\rho_{AB}).
\end{equation}
Once we have $\rho_A$, the entanglement entropy of subsystem $A$ is given by: 
\begin{equation}
S_A = -\text{Tr}(\rho_A \ln \rho_A). 
\end{equation}
If $\mathrm{Tr(\rho_A^2)} = 1$, our subsystem $A$ is in a pure state. If $\mathrm{Tr(\rho_A^2)} < 1$, our subsystem $A$ is in a mixed state \cite{lecture_notes}. 

\begin{figure}[htbp]
  \centering 
  \includegraphics[width=0.3\textwidth]{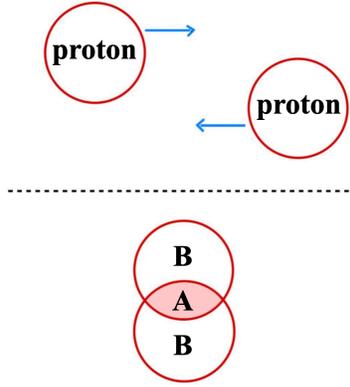}
  \caption{Diagram depicting a proton-proton collision. (Top) Both protons before they collide. (Bottom) The two protons during the collision, where region $A$ is the collision region (region of overlap) and region $B$ is the spectator region (non-overlap region). Regions $A$ and $B$ are entangled.} 
  \label{fig:pp_collision} 
\end{figure}

When two protons collide, both protons are initially in a pure state (see Fig.\thinspace\ref{fig:pp_collision}). 
Once the protons have collided, two regions are present: an overlap (collision) region, $A$, and a non overlap (spectator) region, $B$. In Fig.\thinspace\ref{fig:pp_collision}, $A \cup B$ represents a pure state, since we are considering the proton-proton system as a whole. However, when considering $A$ or $B$ separately, they are each in a mixed state.

\section{Results and Analysis}
\begin{figure}[htbp]
  \centering 
  \includegraphics[width=0.3\textwidth,viewport=0 0 489 436]{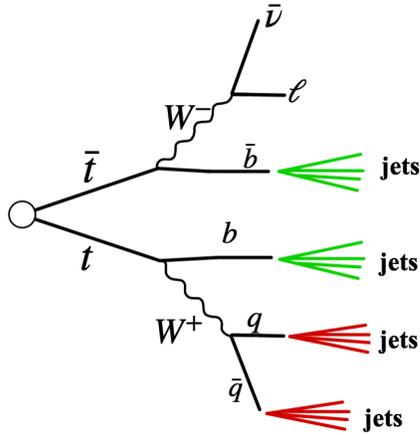}
  \caption{Top anti-top quark decay in the semi-leptonic channel. The resulting W bosons can decay hadronically, resulting in a quark antiquark pair or leptonically, resulting in a lepton and a neutrino \cite{cern_website}.} 
  \label{fig:feynman_diagram} 
\end{figure}

We begin our study by using the transverse momentum distribution of $t\bar{t}$ pair production in the semi-leptonic decay channel, focusing on the hadronic decay products, which is described by:
\begin{equation}
t\bar{t} \rightarrow W^{+}bW^{-}b \rightarrow  q\bar{q}b + \ell\bar{\nu}\bar{b} + \text{jets}. 
\end{equation}
The process described in Eq.\thinspace 7 is depicted in Fig.\thinspace\ref{fig:feynman_diagram}. 
Throughout this letter, the center of mass ($pp$ collision) energy is $\sqrt{s} = 13$ TeV. The following relations for the thermal and hard scattering components of the transverse momentum distribution (Eqs.\thinspace 8 and 11) were originally proposed in Ref.\thinspace\cite{Bylinkin:2010uz} and have been used in other studies \cite{Bylinkin:2014vra, Iskander:2020rkb, Liu, Baker:2017wtt, Weber:2019}. The thermal component of the transverse momentum distribution is given by the following,
\begin{equation} 
\frac{1}{p_T} \frac{d\sigma}{dp_T} = A_{\text{th}} \exp\left(-\frac{m_T}{T_{\text{th}}}\right),
\end{equation} 
where $p_{T}$ is the transverse momentum of the system, $A_{th}$ is a fitting parameter, $m_{T}$ is the transverse mass of the system, and $T_{th}$ is the effective (thermal) temperature parameter. 

The transverse mass is calculated using the following relation, 
\begin{equation}
m_{T}^2 = m^2 + p_{T}^2,
\end{equation}
where $m$ is the mass of the $t\bar{t}$ system, which in this case is the mass of top quark \& anti-top-quark together $\approx 2\times(173 \thinspace\text{GeV}/\text{c}^{2})$.
The effective temperature parameter, which was extracted in Ref.\thinspace\cite{Bylinkin:2014vra}, is given by, 
\begin{equation}
T_{\text{th}} = 0.098 \left(\frac{s}{s_0}\right)^{0.06}\, \text{GeV},
\end{equation} 
where $\sqrt{s_{0}}$ is a normalization constant equal to $\sqrt{s_{0}} = 1$ \text{GeV} and $\sqrt{s}$ is the proton-proton collision energy (which is $\sqrt{s} = 13$ \text{GeV}). 

The hard scattering component of the transverse momentum distribution is given by: 
\begin{equation}
\frac{1}{p_T} \frac{d\sigma}{dp_T} = \frac{A_{\text{hard}}}{\left(1 + \frac{m_{T}^2}{T_{hard}^2 n}\right)^n} . 
\end{equation}
In Eq.\thinspace11, $A_{hard}$ is a fitting parameter, $T_{hard}$ is the hard scale parameter, and $n$ is a scaling factor obtained from the power law fit. The values $m_{T}$, $\sqrt{s_{0}}$ and $\sqrt{s}$ remain unchanged from their previous definitions. 

The hard scale parameter, which was determined in Ref.\thinspace\cite{Bylinkin:2014vra}, is defined as: 
\begin{equation}
T_{\text{hard}} = 0.409 \left(\frac{s}{s_0}\right)^{0.06}\, \text{GeV}.
\end{equation}

\begin{figure}[htbp]
  \centering 
  \includegraphics[width=0.8\textwidth]{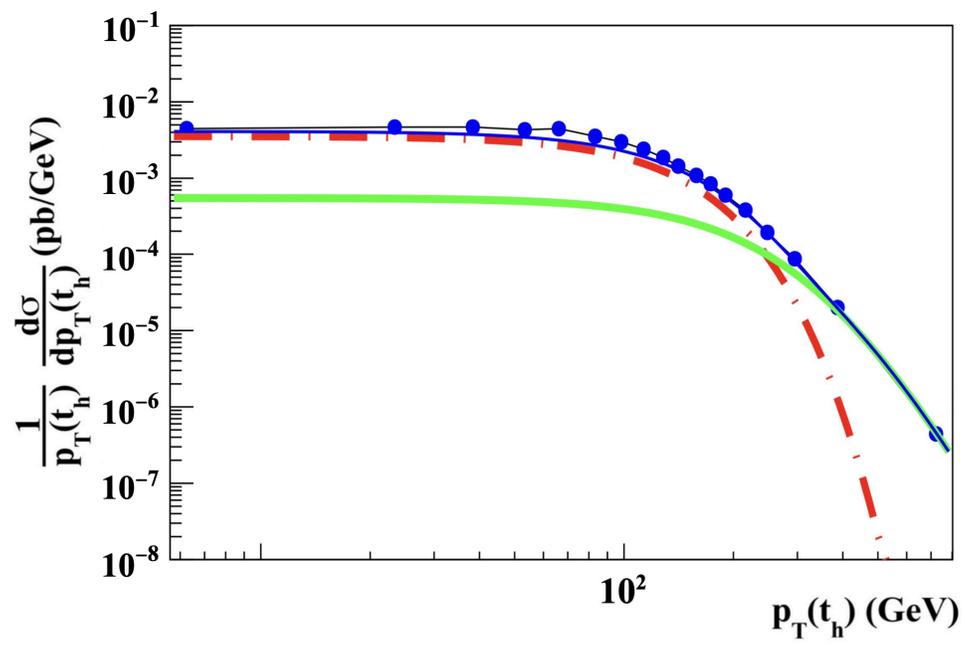}
  \caption{Transverse momentum distribution of top-antitop quark pair production from ATLAS data, with a center of mass energy of 13 TeV and a luminosity of 3.2 fb$^{-1}$. The reduced chi-squared fit value is $\chi^2 / \text{ndf} \approx 24.7/15 = 1.6$. Data is taken from \cite{Fabbri:2019vxz}.} 
  \label{fig:hadronic_ATLAS} 
\end{figure}

The CERN ROOT fitting program and the SciPy \texttt{curve\_fit}  function were used to fit Eqs.\thinspace8 and 11 to $t\bar{t}$ decay data arising from proton-proton collisions at the Large Hadron Collider. The transverse momentum distribution of $t\bar{t}$ production for the hadronic decay products in the semi-leptonic decay channel (ATLAS) with an integrated luminosity of $3.2$ fb$^{-1}$, is depicted in Fig.\thinspace\ref{fig:hadronic_ATLAS}.
As we can see, both a thermal component (red) and a hard scattering component (green) are needed to properly fit the data, which suggests the presence of entanglement. The sum of the two fits, which has a reduced chi-squared value of $\chi^2 / \text{ndf} \approx 1.6$, is the blue curve. The error bars are smaller than the size of the data points. 
\begin{figure}[htbp]
  \centering 
  \includegraphics[width=0.8\textwidth]{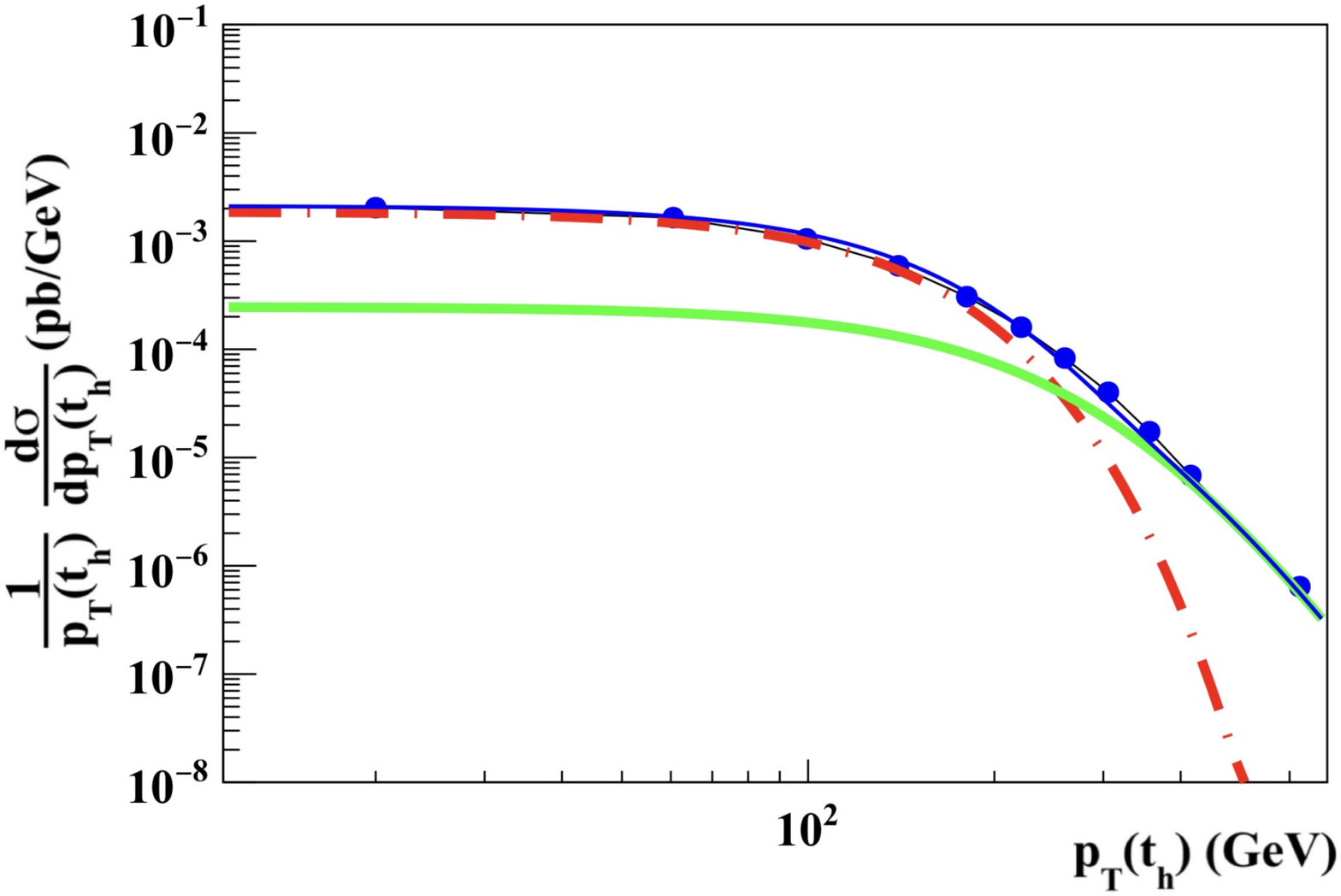}
  \caption{Transverse momentum distribution of top-antitop quark pair production from CMS data, with a center of mass energy of 13 TeV and a luminosity of 35.8 fb$^{-1}$. The reduced chi-squared fit value is $\chi^2 / \text{ndf} \approx 10.3/8 = 1.3$. Data is taken from \cite{Fabbri:2019vxz}.} 
  \label{fig:cms_hadronic} 
\end{figure}
Fig.\thinspace\ref{fig:cms_hadronic} depicts an analogous fit using CMS data, which yielded a reduced chi-squared value of $\chi^2 / \text{ndf} \approx 1.3$. This increase in statistical precision was expected, since the integrated luminosity of the CMS data was $35.8$ fb$^{-1}$, an order of magnitude larger than that of the ATLAS data. As the integrated luminosity increases, the number of recorded events increases as well, which results in a more precise transverse momentum distribution. Again, in Fig.\thinspace\ref{fig:cms_hadronic}, we can see the necessity of having both a thermal and a hard scattering component in the fit.  
\begin{figure}[htbp]
  \centering 
  \includegraphics[width=0.8\textwidth]{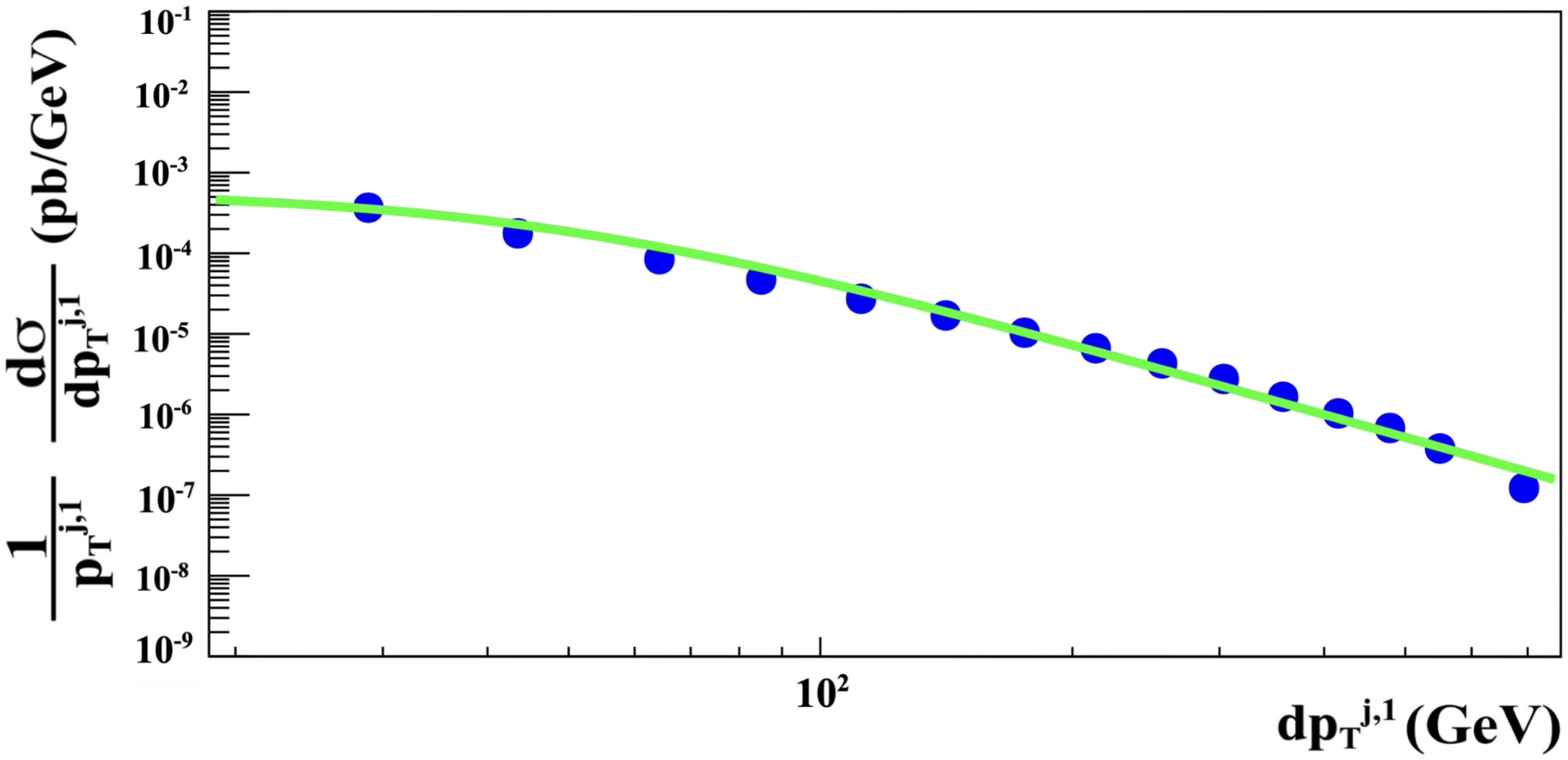}
  \caption{Transverse momentum distribution of top-antitop quark pair production from ATLAS data, with respect to the additional leading jet. Center of mass energy is 13 TeV and luminosity is 139 fb$^{-1}$. The reduced chi-squared fit value is $\chi^2 / \text{ndf} \approx 17.0/13 = 1.3$. Data is taken from \cite{ATLAS:2022xfj}.} 
  \label{fig:one_additional_jet} 
\end{figure} 

Fig.\thinspace\ref{fig:one_additional_jet} depicts the transverse momentum distribution of the additional leading jet. The emergence of the additional leading jet in proton-proton collisions prompts an important question: it does not directly stem from the initial proton-proton state and may lack inherent entanglement with other quantum subsystems. While the initial proton-proton system typically exhibits entanglement between spectator and collision regions, resulting in thermal behaviors observed in variables like the top quark decay jet $p_T$ spectra, the additional jet appears subsequent to the primary collision, driven by secondary QCD radiation and underlying event processes. These processes operate independently of the initial proton system, leading to the absence of any predefined relationship between the spectator and collision regions of the protons. As we can see in Fig.\thinspace\ref{fig:one_additional_jet}, only the hard scattering component is needed to fit the data, implying the absence of entanglement, as predicted. 

One can quantify the presence of a thermal component in the transverse momentum distribution by calculating the ratio between the area under the curve (integral) of the hard scattering component (Eq.\thinspace11) and the area under the curve (integral) of the sum of the fits (Eq.\thinspace8 + Eq.\thinspace11). This ratio, $R$, is defined as,
\begin{equation}
R = \frac{I_p}{I_e + I_p},
\end{equation}
where $I_p$ is the area under hard scattering (power law) portion of the curve and $I_e$ is the area under the thermal (exponential) part of the curve. If there is no thermal component to the fit, $I_e = 0$. When ATLAS data was used, $R$ was calculated to be 0.19 $\pm$ 0.03. Using CMS data yielded a slightly different $R$, which was 0.16 $\pm$ 0.03. When we examined the transverse momentum with respect to the additional leading jet of the system, the $R$ was found to equal one. This  is exactly what we expected, as there was no entanglement present in this case. The calculated $R$ values are consistent with the ratios computed for other $pp$ collisions, as well the $R$ values for charged weak interactions \cite{Iskander:2020rkb}, \cite{Baker:2017wtt}, \cite{Weber:2019}. For the process given by $\bar{\nu_{\mu}} + \text{}^{12}\text{C} \rightarrow \mu^+ + \pi^- + \text{}^{12}\text{C}$, no entanglement was expected since the event was diffractive, which implies that the nucleus as a whole was probed. Therefore, there were no identifiable collision and spectator regions which could be entangled with one another. This parallels our result for the transverse momentum distribution of the additional leading jet, since in this case, distinguishable collision and spectator regions were also absent. Table 1 summarizes the results from previous literature as well as our new results. 

\begin{table}[h]
  \centering
\begin{tabular}{|c|c|c|}
  \hline
  $R$ & Process & Reference \\
  \hline
  0.16 $\pm$ 0.05 & $pp \rightarrow$ charged hadrons & \cite{Baker:2017wtt}, \cite{Weber:2019}\\
  0.15 $\pm$ 0.05 & $pp \rightarrow \text{H} \rightarrow \gamma\gamma$ & \cite{Baker:2017wtt}, \cite{Weber:2019} \\
  0.23 $\pm$ 0.05 & $pp \rightarrow \text{H} \rightarrow 4l(e, \mu)$ & \cite{Baker:2017wtt}, \cite{Weber:2019} \\
  1.00 $\pm$ 0.02 & $pp(\gamma\gamma) \rightarrow (\mu\mu)\text{X'} \text{X''}$ & \cite{Baker:2017wtt}, \cite{Weber:2019} \\
  0.13 $\pm$ 0.03 & $\bar{\nu_{\mu}} + N \rightarrow \mu^+ + \pi^0 + X $ & \cite{Iskander:2020rkb} \\
  1.00 $\pm$ 0.05 & $\bar{\nu_{\mu}} + \text{}^{12}\text{C} \rightarrow \mu^+ + \pi^- + \text{}^{12}\text{C} $ & \cite{Iskander:2020rkb} \\
  0.19 $\pm$ 0.03 & $ pp \rightarrow t\bar{t} \rightarrow WbWb$ (ATLAS) & current work \\
  0.16 $\pm$ 0.03 & $ pp \rightarrow t\bar{t} \rightarrow WbWb$ (CMS) & current work \\
  1.00 $\pm$ 0.05 & $ pp \rightarrow t\bar{t} \rightarrow WbWb \rightarrow$ jets (ATLAS)& current work \\
  \hline
\end{tabular}
\caption{$R$ values from prior studies and our current work.}
\end{table}

\section{Conclusion}
In this letter, we have extended upon the ideas  in Refs.\thinspace\cite{Bylinkin:2014vra, Iskander:2020rkb, Liu, Baker:2017wtt, Weber:2019} to show that even in $t\bar{t}$ collisions, the thermal component of the transverse momentum distribution can be attributed to entanglement between different parts of the wave functions of the colliding particles (in this case, protons). In Ref.\thinspace\cite{Duan:2021kmf}, Duan discusses this idea further, introducing a term called “entropy of ignorance.” In his work, Duan agrees that in proton-proton collisions, there is entanglement between the collision and spectator regions of the proton. Since an experiment can only measure the collision region, we lack information about the spectator region. This lack of information is called the “entropy of ignorance.” 

Studies of entanglement in $t\bar{t}$ collisions can also be used to investigate possibilities of physics beyond the standard model. In Ref.\thinspace\cite{Afik:2022dgh}, a term called quantum discord is discussed, which is a fundamental quantity that measures the “quantumness of correlations" \cite{Brodutch}. If the quantum discord is asymmetric, this can hint at the presence of CP violation. It would be interesting to apply these ideas to new experimental measurements of $t\bar{t}$ pair production or to other types of particle collisions. 

\section*{Acknowledgements}
The authors gratefully acknowledge funding support from the Department of Energy Office of Science Award DE-FG02-92ER40704.

\end{document}